\documentclass{osa-article}

\journal{osac}

\articletype{Research Article}
\usepackage{graphics}
\usepackage{subfigure}
\usepackage{lineno}

\usepackage{float}
\begin{document}

\title{Generation of large-scale continuous-variable cluster states multiplexed both in time and frequency domains}

\author{peilin du,\authormark{1,2} yu wang,\authormark{1,2} kui liu,\authormark{1,2,3,4} rongguo yang\authormark{1,2,4,5} and jing zhang,\authormark{1,2,4,6}}

\address{\authormark{1}State Key Laboratory of Quantum Optics and Quantum Optics Devices, Shanxi University, Taiyuan 030006, China\\
\authormark{2}College of Physics and Electronic Engineering, Shanxi University,Taiyuan 030006, China\\
\authormark{3}Institute of Opto-Electronics, Shanxi University, Taiyuan 030006, China\\
\authormark{4}Collaborative Innovation Center of Extreme Optics, Shanxi University, Taiyuan, 030006, China\\
\authormark{5}yrg@sxu.edu.cn\\
\authormark{6}zjj@sxu.edu.cn}



\begin{abstract}
Large-scale continuous variable (CV) cluster state is necessary in quantum information processing based on measurement-based quantum computing (MBQC). Specially, generating large-scale CV cluster state multiplexed in time domain is easier to implement and has strong scalability in experiment. Here one-dimensional (1D) large-scale dual-rail CV cluster states multiplexed both in time and frequency domains are parallelly generated, which can be further extended to a three-dimensional (3D) CV cluster state by combining two time-delay NOPA systems with beam-splitters. It is shown that the number of parallel arrays depends on the corresponding frequency comb lines, the partite number of each array can be very large (million), and the scale of the 3D cluster state can be ultra-large. In addition, the concrete quantum computing schemes of applying the generated 1D and 3D cluster states are also demonstrated. Our schemes may pave the way for fault-tolerant and topologically protected MBQC in hybrid domains, by further combing with efficient coding and quantum error correction.
\end{abstract}

\section{Introduction}
\label{sec:1}

Based on unique quantum properties (such as superposition and entanglement), quantum computing (QC) can provide effective solutions for some special problems that are out of reach for classical computers\cite{Nielsen(2000),feynman(1982),horodecki(2009),Bennett(2000)}. Measurement-based quantum computing (MBQC), as one of the experimentally feasible quantum computing models, provides the ability to perform QC using projection measurements, and requires specially prepared cluster states\cite{raussendorf(2001),Briegel(2001)prl}. Cluster state is a kind of multi-mode entangled state, in which interactions only exist between adjacent modes. Multi-partite cluster state with partite number $N>4$ has a stronger ability of entanglement retention, compared with GHZ state\cite{ZhangJing(2006),MGu(2009),Menicucci(2006)}. Therefore, researches on large-scale continuous variable (CV) cluster state have attracted more attentions recent years.

As is known, the CV cluster state can be generated by spatial separation\cite{VANLOOCK(2007)} or frequency/time multiplexing\cite{menicucci(2007),alexander(2018)}. By using several OPO/OPAs and a beam-splitter network, the four- and six- and eight-partite CV cluster states\cite{yukawa(2008),su(2007),su(2013),suxiaolong(2012)} were generated in spatial separation way, which lacks scalability because huger experimental setup and more complex phase-tuning techniques are required for further more entanglement partites. More entanglement partites, such as fifteen 4-partite, two 30-partite, and one 60-partite CV cluster states\cite{pysher(2011),chenmoren(2014)} were realized in frequency multiplexing way, which integrates many single-mode OPOs into a multi-mode OPO to replace the beam-splitter network and makes the experimental setup more compact. In addition, one-dimensional CV cluster state of  $10^{6}$ partites\cite{yokoyama(2013),yoshikawa(2016)}, and two-dimensional square lattice CV cluster state of $10^{3}$ partites\cite{asavanant(2019),larsen(2019)} (as a universal resource for MBQC) were obtained in time multiplexing way, which has better scalability and is easier to realize in experiment, compared with the frequency multiplexing way (limited by the bandwidth of the cavity).

Furthermore, one can utilize two or more methods mentioned above to get large-scale CV cluster states of rich structures. Large-scale bi-layer square lattice CV cluster state was investigated by time and frequency multiplexing\cite{alexander(2016),wu(2020)}. Spatio-temporal CV cluster states of different structures were demonstrated by combining optical- and spatial-mode multiplexing\cite{zhangJing(2017),yangrg(2020)}. In this paper, we consider both time and frequency multiplexing and propose a scheme to parallel generate large-scale CV cluster states of zig-zag structure, which has strong scalability and can be further extended to generate a three-dimensional (3D) cluster state. 

\section{Generation of 1D cluster states}
\label{sec:2}

As is shown in Fig.1(a), the system mainly consists of a two-sided cavity (with a nonlinear crystal in it), an optical fiber delay device, and a beam-splitter. A pump beam passes through a type II phase-matching nonlinear crystal (second-order nonlinear coefficient is $\xi$) in the cavity and two entangled down-concerted beams (signal and idler, with orthogonal polarization) are generated by the so-called nondegenerate optical parametric amplification (NOPA) process operating below threshold $\left(\sigma<1 \right)$ .  Thus, two optical frequency combs (signal and idler) whose longitudinal eigenmodes are separated by free spectral range (FSR)  can also be generated. Then the signal beam and the time-delayed idler beam are coupled on the beam-splitter and the output states can be obtained.
\begin{figure}[htbp]
\centering
\begin{subfigure}{
\centering
\includegraphics[height=0.08\textheight,width=0.8\linewidth]{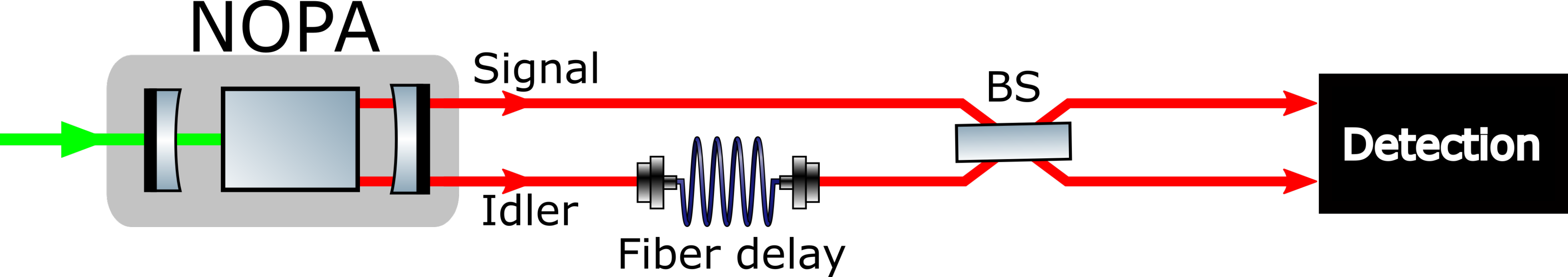}}
\end{subfigure}
\centerline{(a)}
\begin{subfigure}{
\centering
\includegraphics[height=0.16\textheight,width=0.99\linewidth]{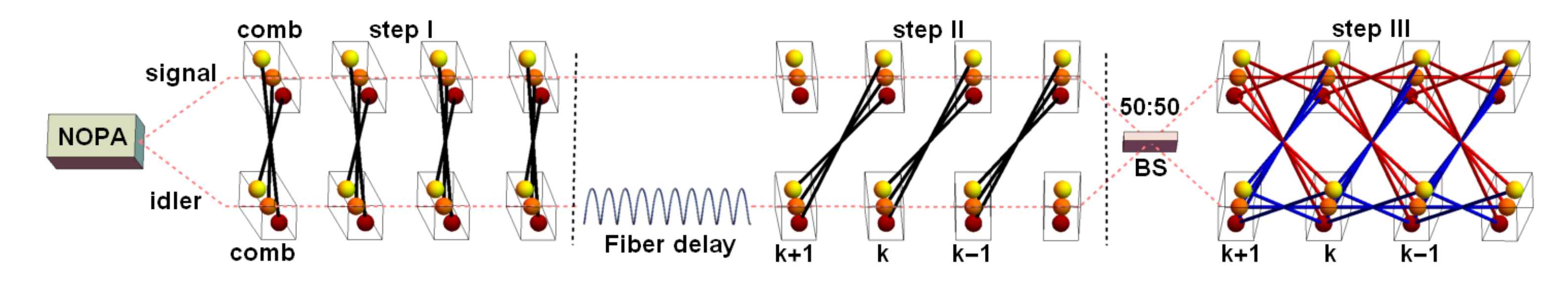}}
\end{subfigure}
\centerline{(b)}
\caption{(a) The schematic diagram  of generating dual-rail cluster states. (b) The detailed physical process. The transparent cuboids represent the corresponding optical frequency combs. Different colors (orange, dark-red, yellow) of the small balls stand for different frequencies. Different colors (black, red and blue) of the lines connecting the small balls represent different entanglement weights.}
\label{fig1}
\end{figure}

The pump field of frequency $\omega_{p}=2\omega_{0}$ can be down-converted into signal and idler fields of frequencies $ \omega_{s,i}=\omega_{0}\pm n \Delta$, $\Delta$ represents the FSR of the cavity and $n=0,1,2.... $. Nonlinear interaction must satisfy the conservation of energy $\left(\omega_{p} =\omega_{s}+\omega_{i}\right)$ and momentum $\left(\vec{k}_{p} =\vec{k}_{s}+\vec{k}_{i}\right)$. Each pair of modes connected by the black line in Fig.1(b) come from the same optical down-conversion process, which correspond to an EPR state. The interaction Hamiltonian of step \uppercase\expandafter{\romannumeral 1} is: 
\begin{equation}
H=i\hbar \xi \sum\limits_{i}{{\left\{ {{G}_{(\omega_{0} \pm i\Delta ),(\omega_{0} \mp i\Delta)}}{{b}_{\left(2\omega_{0} \right)}}a_{(\omega_{0} \pm i\Delta)}^{\dagger }a_{(\omega_{0} \mp i\Delta)}^{\dagger } \right\}}}+H.c.,
\end{equation}
where $G$ matrix specifies the (linearized) Hamiltonian for the NOPA that acts on the vacuum of the cavity modes to generate the Hamiltonian graph ($H$ graph) state and denotes the adjacency matrix of this state\cite{menicucci(2007),menicucci(2011),MenicucciGraphical(2011)}, whose corresponding matrix element is $1$ when the parametric process exists, and $0$ otherwise. ${b}_{\left(2\omega_{0}\right)}$ represents the annihilation operator of pumping mode with frequency $2\omega_{0}$, which can be considered as a classical number, since the pump beam is single-pass and not correlated with the down-converted beams. $a_{(\omega_{0} \pm i\Delta)}^{\dagger }$ and $a_{(\omega_{0} \mp i\Delta)}^{\dagger }$ are the creation operators of the signal and idler fields, respectively. Thus, the multiplication item expresses the corresponding down-conversion process. 

Here, only three frequencies $ \omega_{0}$ (orange), $ \omega_{0}+\Delta$ (yellow) and $ \omega_{0}-\Delta$ (dark-red) of its optical frequency comb (OFC) are considered, for simplicity. The continuous-wave signal and idler fields can be divided into time bins of time period T, where 1/T is narrower than the bandwidth of the NOPA cavity. In step I of Fig.1(b), a series of OFCs separated by time interval T can be deterministically created, where each wave packet (transparent cuboid) of signal and idler fields in each time bin represents a mutually independent OFC. After the fiber delay, the idler comb of k time are synchronized with the signal comb of k+1 time, as is shown in step II of Fig.1(b). By combining the staggered combs on the balanced beam-splitter, each comb interacts with the former and later combs, which result in the connection of all neighbour combs, thereby the cluster state in step III of Fig.1(b) is generated.

The parallelly generated zig-zag-structured dual-rail cluster states are shown more clearly in Fig.2, and the connections are always between $\omega_{0}\pm i\Delta$ and $\omega_{0}\mp i\Delta$ (only i=0,1 here). Thus, the more modes the comb has, the more dual-rail cluster states can be generated in parallel. Note that the bandwidth of the OFC is mainly affected by the nonlinear process in NOPA. The down-conversion bandwidth in a bulk nonlinear material of length $L_{crystal}$ can be estimated to be $10c/L_{crystal}$ (about 10 THz), and the FSR of a cavity with length $L_{cavity}$ is $c/2L_{cavity}$ (about 1 GHz), $c$ is the speed of light. Thus, the number of zig-zag-structured dual-rail cluster states is approximately $10^4$ in the experiment\cite{pysher(2011),chenmoren(2014)}.
\begin{figure}[H]
\centering
\includegraphics[width=1\linewidth]{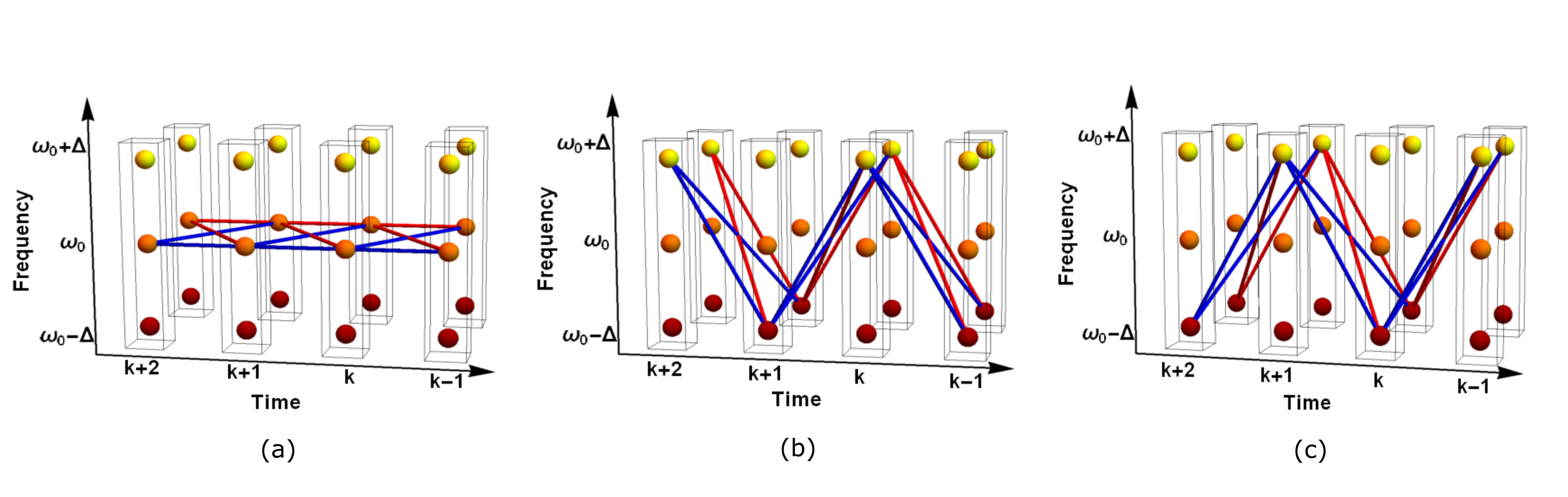}
\caption{Parallelly generated dual-rail cluster states. (a) dual-rail state consisting of frequency $\omega_{0}$ . (b)-(c) dual-rail states connecting frequencies $\omega_{0}\pm\Delta$ and $\omega_{0}\mp\Delta$. Red (blue) lines represent entanglement weights (1/2) $/$ (-1/2).} 
\label{ruselt}
\end{figure}

\section{Generation of 3D cluster state}
\label{sec:3}

To achieve the universal MBQC, the dimension of the cluster state must be at least two. Based on the above system, 3D cluster state can be generated by the divide-and-conquer method as shown in Fig.3.
\begin{figure}[h]{
\centering
\includegraphics[height=0.2\textheight,width=1\linewidth]{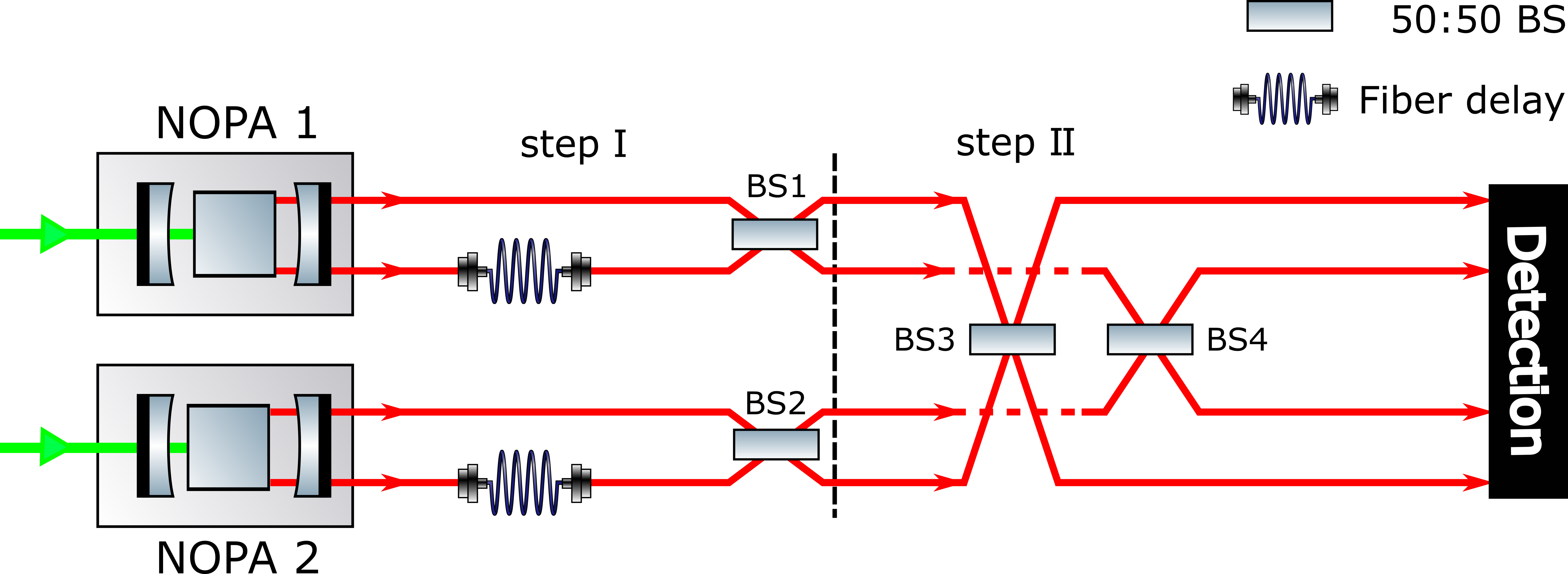}}
\caption{ The schematic diagram of generating 3D cluster state.}	\label{fig:2d}
\end{figure}

First, two identical "NOPA + delay" systems with the different pump frequencies $(2\omega_{0}+\Delta$ and $2\omega_{0}-\Delta)$ are used to generate two sets of different dual-rail cluster states in step \uppercase\expandafter{\romannumeral 1} of Fig.3. Then the signal field of one system and the idler field of the other system are coupled on the balanced beam-splitters (BS3 and BS4). After coupling, all dual-rail cluster states can be entirely entangled, which forms an entanglement structure shown in Fig.4(a). Note, if we choose the same pump frequency, it will produce two sets of identical dual-rail cluster states. After coupling they will still be two sets of identical dual-rail cluster states and will not form more complex entangled structures. 

As is shown in Fig.4(a), each node (red and aqua ball) in the structure contains two modes (signal and idler modes at the same time) with the same frequency, and each edge (black line) represents a dual-rail structured connection. The 3D cluster state is bipartite, or bicolorable, which means that all modes can be expressed by two  colors and there are no interactions among nodes with the same color.  Fig.4(b) gives a detailed diagram of a unit (gray cube) of the whole structure shown in Fig.4(a), where the subscript represents the time, and superscript $1$ and $2$ represent the NOPA1 and NOPA2, respectively. Its unit structure contains four modes of frequencies $\omega_0 \pm a\Delta$ at k time, $\omega_0 \pm a\Delta$ at k+2 time, $\omega_0\pm b\Delta $ at k+1 time and  $\omega_0 \pm c\Delta$ at k+1 time, which form a "cube" in Fig.4(a). 
\begin{figure}[H]{
\centering
\includegraphics[height=0.25\textheight,width=1\linewidth]{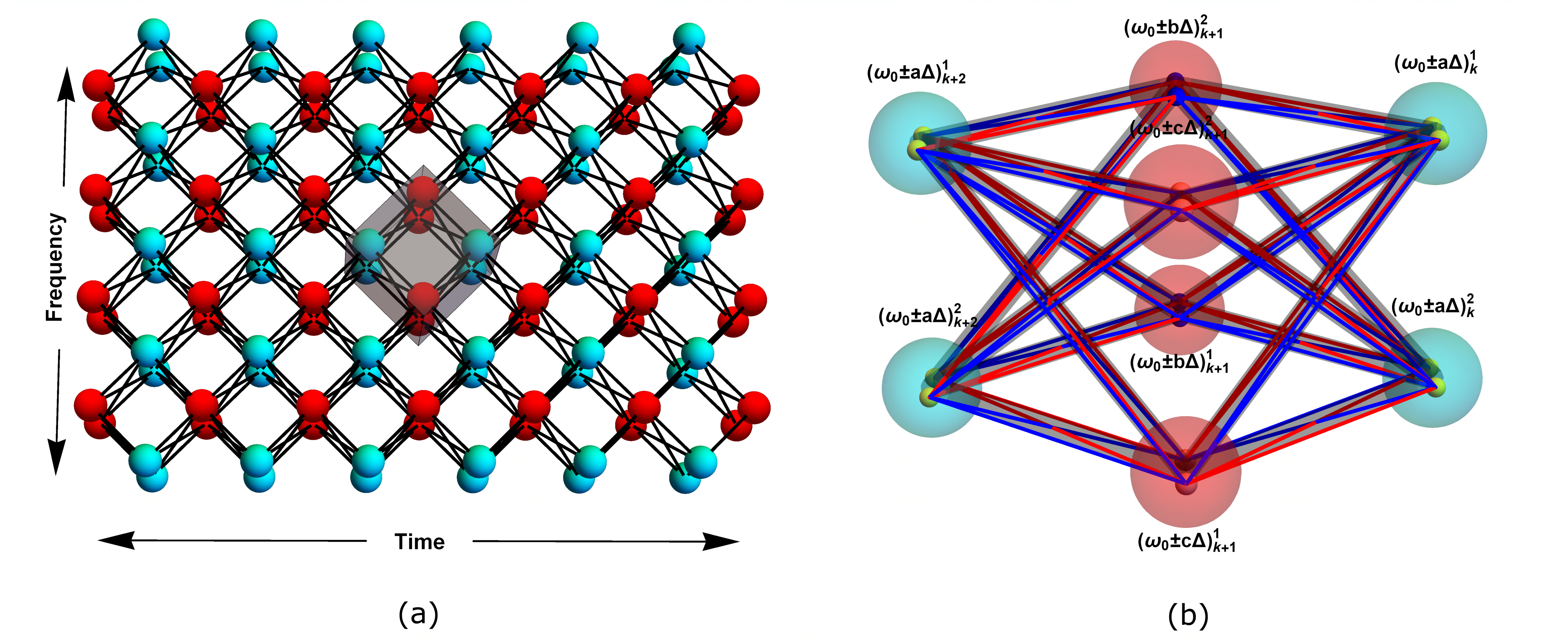}}
\caption{Whole structure (a) and detailed structure (b) of the generated 3D cluster state. In Fig.4(b), the red/blue lines represent the entanglement weights of (1/4)/(-1/4).}	\label{fig:2}
\end{figure}

\section{Nullifier and full-inseparability criterion}
\label{sec:4}

When the $G$ matrix in equation (1) is self-inverse and bipartite, the complex adjacency matrix $Z$ of Gaussian pure state can be written as\cite{menicucci(2011),MenicucciGraphical(2011)}:\begin{equation}
    Z=i\cosh\left ( 2r\right )I-i\sinh\left ( 2r\right )G,
    	\label{z=cosh+sinh}  
\end{equation}
where $I$ represents an identity matrix. This $H$ graph state can be transformed into a cluster state by applying a $\frac{\pi }{2}$ phase-space rotation on half of its modes. Since this $\frac{\pi }{2}$ rotation can be absorbed into the measurement basis when measuring each mode of the state, a self-inverse bipartite $H$ graph state (for example, the dual-rail and square-lattice structured states) can be considered as a cluster state. The approximate nullifiers for this cluster state can be obtained as:
\begin{eqnarray}
\begin{aligned}
\left(\hat{X}- G \hat{X}\right) \left| \Psi _Z\right\rangle \xrightarrow{r\longrightarrow\infty} 0,  \\
\left(\hat{P}+G \hat{P}\right) \left| \Psi _Z\right\rangle\xrightarrow{r\longrightarrow\infty} 0,      
\end{aligned}
\end{eqnarray} 
where $\hat{X}$ and $\hat{P}$ are column matrices, and represent the quadrature amplitude and quadrature phase, respectively. $\left| \Psi _Z\right\rangle $ is the created cluster state.

Here, the dual-rail cluster states generated in step \uppercase\expandafter{\romannumeral3} of Fig.1(b) can be expressed as:
\begin{equation}
\left( 	\begin{array}{c}
\hat{a}^{\left( 1\right) } _{s,k}\\
\hat{a}^{\left( 1\right) } _{i,k}
\end{array}\right)=\frac{1}{\sqrt{2}}\left( 	\begin{array}{c}
\hat{a}^{\left( 0\right) } _{s,k}-\hat{a}^{\left( 0\right) } _{i,k}-	\hat{a}^{\left( 0\right) } _{s,k-1}-\hat{a}^{\left( 0\right) } _{i,k-1}\\
\hat{a}^{\left( 0\right) } _{s,k}-\hat{a}^{\left( 0\right) } _{i,k}+\hat{a}^{\left( 0\right) } _{s,k-1}+\hat{a}^{\left( 0\right) } _{i,k-1}
\end{array}\right),
\end{equation}
where the superscript (0) and (1) of the annihilation operators correspond to vacuum state and dual-rail state. The subscripts correspond to down-conversion field and the time, respectively. Then, we introduce the nullifiers to characterize the generated cluster states which correspond to the stabilizer for cluster states with discrete variables in the case of the infinite squeezing and are used to verify the generation of cluster states. The nullifiers of the dual-rail cluster states can be obtained by calculating the relations of the annihilation operators\cite{yokoyama(2013),yoshikawa(2016),fukui(2020)},
\begin{eqnarray}
\begin{aligned}
\hat{n}_{k}^x=\hat{x}^s_{k}+\hat{x}^i_{k}-\hat{x}^s_{k+1}+\hat{x}^i_{k+1},\\
\hat{n}_{k}^p=-\hat{p}^s_{k}-\hat{p}^i_{k}-\hat{p}^s_{k+1}+\hat{p}^i_{k+1},
\end{aligned}
\end{eqnarray}where the superscript and subscript of the quadrature operators represent the down-conversion field and the time, respectively. 

Subsequently, we use the Van Loock-Furusawa (VLF) full inseparable criterion. For a minimum unit of the dual-rail cluster states, there are four basic modes and seven possible patterns, based on which the following inequalities as a sufficient condition due to full inseparability can be obtained as:\begin{equation}
\left \langle \left (\hat{n}_{k}^x\right )^2\right \rangle<\hbar\qquad       \left \langle \left (\hat{n}_{k}^p \right )^2\right \rangle<\hbar\qquad for\  all\  k,
\end{equation}
where $\hbar=1 $, and variances of nullifiers $\hat{n}_{k}^x $ and $\hat{n}_{k}^p$ can be  calculated,
\begin{equation}
\begin{aligned}
\left \langle \left ( \hat{x}^s_{k}+\hat{x}^i_{k}-\hat{x}^s_{k+1}+\hat{x}^i_{k+1}\right )^2\right\rangle=2e^{-2r_s}< 1,\\
\left \langle \left ( -\hat{p}^s_{k}-\hat{p}^i_{k}-\hat{p}^s_{k+1}+\hat{p}^i_{k+1}\right )^2\right\rangle=2e^{-2r_i}< 1,
\end{aligned} 
\end{equation}which determine the required minimum squeezing for generating 1D cluster states. When the squeezing is larger than $-3dB$, the sufficient condition for inseparability will be satisfied, as is shown in Fig.5.
\begin{figure}
\begin{minipage}{0.49\linewidth}
\vspace{3pt}
\centerline{\includegraphics[width=0.8\textwidth]{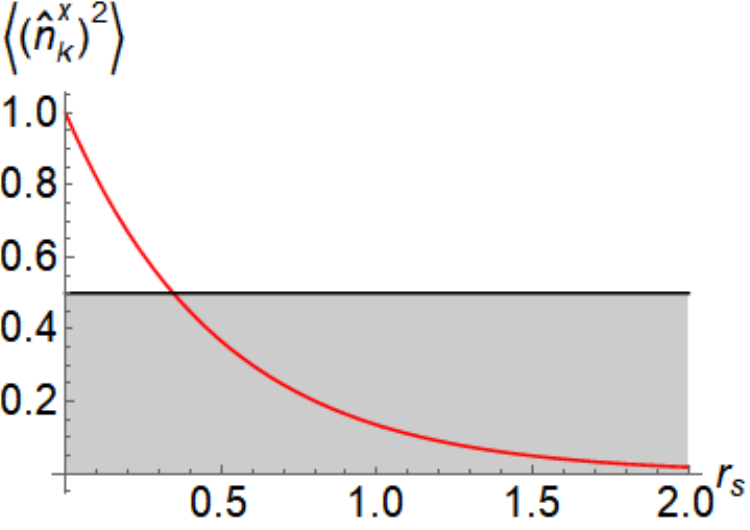}}
\centerline{(a)}
\end{minipage}
\begin{minipage}{0.49\linewidth}
\vspace{3pt}
\centerline{\includegraphics[width=0.8\textwidth]{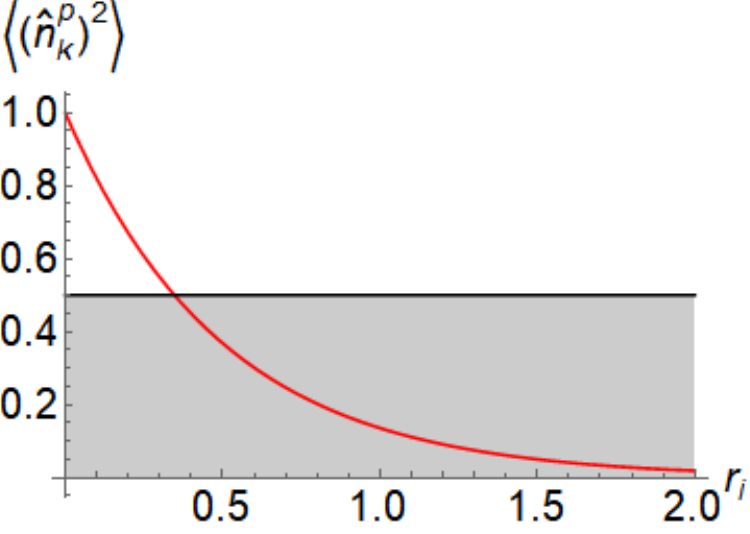}}
\centerline{(b)}
\end{minipage}
\caption{The nullifier of quadrature $\hat{x}$ (a) and $\hat{p}$ (b) versus the squeezing parameter, for the generated 1D cluster state. The criterion of inseparability is satisfied under the black line, and the squeezing at the intersection is about $-3dB$.}
\end{figure}
  
In addition, considering these annihilation operators for 3D cluster state, the two different dual-rail cluster states generated in step I of Fig.3 can be expressed as
\begin{equation}\label{key}
\left( 	\begin{array}{c}
\hat{a}^{s_{1},\left( 1\right) } _{\omega _{a},2k}\\
\hat{a}^{i_{1},\left( 1\right) } _{\omega _{a},2k}
\end{array}\right)=\frac{1}{\sqrt{2}}\left( 	\begin{array}{c}
\hat{a}^{s_{1},\left( 0\right) } _{\omega _{a},2k}-\hat{a}^{i_{1},\left( 0\right) } _{\omega _{a},2k}-\hat{a}^{s_{1},\left( 0\right) } _{\omega _{b},2k-1}-\hat{a}^{i_{1},\left( 0\right) } _{\omega _{b},2k-1}\\
\hat{a}^{s_{1},\left( 0\right) } _{\omega _{a},2k}-\hat{a}^{i_{1},\left( 0\right) } _{\omega _{a},2k}+\hat{a}^{s_{1},\left( 0\right) } _{\omega _{b},2k-1}+\hat{a}^{i_{1},\left( 0\right) } _{\omega _{b},2k-1}\\
\end{array}\right),
\end{equation}
\begin{equation}
\left( 	\begin{array}{c}
\hat{a}^{s_{2},\left( 1\right) } _{\omega _{a},2k}\\
\hat{a}^{i_{2},\left( 1\right) } _{\omega _{a},2k}
\end{array}\right)=\frac{1}{\sqrt{2}}\left( 	\begin{array}{c}
\hat{a}^{s_{2},\left( 0\right) } _{\omega _{a},2k}-\hat{a}^{i_{2},\left( 0\right) } _{\omega _{a},2k}-	\hat{a}^{s_{2},\left( 0\right) } _{\omega _{c},2k-1}-\hat{a}^{i_{2},\left( 0\right) } _{\omega _{c},2k-1}\\
\hat{a}^{s_{2},\left( 0\right) } _{\omega _{a},2k}-\hat{a}^{i_{2},\left( 0\right) } _{\omega _{a},2k}+\hat{a}^{s_{2},\left( 0\right) } _{\omega _{c},2k-1}+\hat{a}^{i_{2},\left( 0\right) } _{\omega _{c},2k-1}
\end{array}\right),
\end{equation}
where $\hat{a}^{D_{1}}_{\omega _{n},k}$ denotes the annihilation operator of the down-converted field $D$ with frequency $\omega _{n}$ from the NOPA1 at k time. $\omega_a = \omega_0 \pm a\Delta$, $\omega_b = \omega_0 \pm b\Delta$, $\omega_c = \omega_0 \pm c\Delta$, corresponding to the OFC's frequencies $\omega_{n}=\omega_{0}\pm n \Delta $. $a,b,c\in n$, and satisfy $a+b=1$ and $a+c=-1$. Then, their signal and idler fields are coupled on BS3 and BS4 in step II of Fig.3. The beam-splitter transformation can be defined as the following unitary matrix acting on annihilation operators: 
\begin{equation}
\hat{U}_{BS} \begin{pmatrix}
a_{\omega_{n,2k }}^{D_{1}} \\
a_{\omega_{n,2k }}^{D_{2}}
\end{pmatrix}\hat{U}_{BS}^{\dagger } =\frac{1}{\sqrt{2}}\begin{pmatrix}
  1&-1\\
  1&1
\end{pmatrix}\begin{pmatrix}
a_{\omega_{n,2k}}^{D_{1}} \\
a_{\omega_{n,2k }}^{D_{2}}
\end{pmatrix},
\end{equation}
and the output 3D cluster state can be described as:
\begin{align}
&\left( 	\begin{array}{c}
\hat{a}^{s_{1},\left( out\right) } _{\omega _{n},2k}\\
\hat{a}^{i_{2},\left( out\right) } _{\omega _{n},2k}
\end{array}\right)=\frac{1}{\sqrt{2}}\left( \begin{array}{cc}
1 & -1 \\
1 & 1
\end{array}\right) \left( 	\begin{array}{c}
\hat{a}^{s_{1},\left( 1\right) } _{\omega _{n},2k}\\
\hat{a}^{i_{2},\left( 1\right) } _{\omega _{n},2k}
\end{array}\right)\nonumber\\& =\frac{1}{2}\left( 	\begin{array}{c}
\hat{a}^{s_{1},\left(0\right) } _{\omega _{n},2k}-\hat{a}^{i_{1},\left(0\right) } _{\omega _{n},2k}-\hat{a}^{s_{1},\left(0\right) } _{\omega _{1-n},2k-1}-\hat{a}^{i_{1},\left(0\right) } _{\omega _{1-n},2k-1}-\hat{a}^{s_{2},\left(0\right) } _{\omega _{n},2k}+\hat{a}^{i_{2},\left(0\right) } _{\omega _{n},2k}-\hat{a}^{s_{2},\left( 0\right) } _{\omega _{-1-n},2k-1}-\hat{a}^{i_{2},\left(0\right) } _{\omega _{-1-n},2k-1}\\
\hat{a}^{s_{1},\left(0\right) } _{\omega _{n},2k}-\hat{a}^{i_{1},\left(0\right) } _{\omega _{n},2k}-\hat{a}^{s_{1},\left(0\right) } _{\omega _{1-n},2k-1}-\hat{a}^{i_{1},\left( 0\right) } _{\omega _{1-n},2k-1}+\hat{a}^{s_{2},\left(0\right) } _{\omega _{n},2k}-\hat{a}^{i_{2},\left(0\right) } _{\omega _{n},2k}+\hat{a}^{s_{2},\left(0\right) } _{\omega _{-1-n},2k-1}+\hat{a}^{i_{2},\left(0\right) } _{\omega _{-1-n},2k-1}
\end{array}\right),\\
&\left( 	\begin{array}{c}
\hat{a}^{i_{1},\left( out\right) } _{\omega _{n},2k}\\
\hat{a}^{s_{2},\left( out\right) } _{\omega _{n},2k}
\end{array}\right)=\frac{1}{\sqrt{2}}\left( \begin{array}{cc}
1 & -1 \\
1 & 1
\end{array}\right) \left( 	\begin{array}{c}
\hat{a}^{i_{1},\left( 1\right) } _{\omega _{n},2k}\\
\hat{a}^{s_{2},\left( 1\right) } _{\omega _{n},2k}
\end{array}\right)\nonumber\\& =\frac{1}{2}\left( 	\begin{array}{c}
\hat{a}^{s_{1},\left(0\right) } _{\omega _{n},2k}-\hat{a}^{i_{1},\left(0\right) } _{\omega _{n},2k}+\hat{a}^{s_{1},\left(0\right) } _{\omega _{1-n},2k-1}+\hat{a}^{i_{1},\left(0\right) } _{\omega _{1-n},2k-1}-\hat{a}^{s_{2},\left(0\right) } _{\omega _{n},2k}+\hat{a}^{i_{2},\left(0\right) } _{\omega _{n},2k}+\hat{a}^{s_{2},\left( 0\right) } _{\omega _{-1-n},2k-1}+\hat{a}^{i_{2},\left(0\right) } _{\omega _{-1-n},2k-1}\\
\hat{a}^{s_{1},\left(0\right) } _{\omega _{n},2k}-\hat{a}^{i_{1},\left(0\right) } _{\omega _{n},2k}+\hat{a}^{s_{1},\left(0\right) } _{\omega _{1-n},2k-1}+\hat{a}^{i_{1},\left( 0\right) } _{\omega _{1-n},2k-1}+\hat{a}^{s_{2},\left(0\right) } _{\omega _{n},2k}-\hat{a}^{i_{2},\left(0\right) } _{\omega _{n},2k}-\hat{a}^{s_{2},\left(0\right) } _{\omega _{-1-n},2k-1}-\hat{a}^{i_{2},\left(0\right) } _{\omega _{-1-n},2k-1}
\end{array}\right).
\end{align}
A beam-splitter transformation is associated with the corresponding transformation of nullifiers, $\hat{N}\longrightarrow \hat{U}\hat{N}\hat{U^{\dagger}}$. Therefore, the nullifiers of 3D cluster state can be obtained as:\begin{equation}
\begin{aligned}
\hat{n}_{2k}^{x1}&=\hat{x}_{2k}^{s_{1},\omega _{a} }+\hat{x}_{2k}^{i_{1},\omega _{a} }+\hat{x}_{2k}^{s_{2},\omega _{a} }+\hat{x}_{2k}^{i_{2},\omega _{a} }-\hat{x}_{2k+1}^{s_{1},\omega _{b} }+\hat{x}_{2k+1}^{i_{1},\omega _{b} }-\hat{x}_{2k+1}^{s_{2},\omega _{b} }+\hat{x}_{2k+1}^{i_{2},\omega _{b} },\\
\hat{n}_{2k}^{p1}&=-\hat{p}_{2k}^{s_{1},\omega _{a} }-\hat{p}_{2k}^{i_{1},\omega _{a} }-\hat{p}_{2k}^{s_{2},\omega _{a} }-\hat{p}_{2k}^{i_{2},\omega _{a} }-\hat{p}_{2k+1}^{s_{1},\omega _{b} }+\hat{p}_{2k+1}^{i_{1},\omega _{b} }-\hat{p}_{2k+1}^{s_{2},\omega _{b} }+\hat{p}_{2k+1}^{i_{2},\omega _{b} },\\
\hat{n}_{2k}^{x2}&=-\hat{x}_{2k}^{s_{1},\omega _{a} }-\hat{x}_{2k}^{i_{1},\omega _{a} }+\hat{x}_{2k}^{s_{2},\omega _{a} }+\hat{x}_{2k}^{i_{2},\omega _{a} }-\hat{x}_{2k+1}^{s_{1},\omega _{c} }+\hat{x}_{2k+1}^{i_{1},\omega _{c} }+\hat{x}_{2k+1}^{s_{2},\omega _{c} }-\hat{x}_{2k+1}^{i_{2},\omega _{c} },\\
\hat{n}_{2k}^{p2}&=\hat{p}_{2k}^{s_{1},\omega _{a} }+\hat{p}_{2k}^{i_{1},\omega _{a} }-\hat{p}_{2k}^{s_{2},\omega _{a} }-\hat{p}_{2k}^{i_{2},\omega _{a} }-\hat{p}_{2k+1}^{s_{1},\omega _{c} }+\hat{p}_{2k+1}^{i_{1},\omega _{c} }+\hat{p}_{2k+1}^{s_{2},\omega _{c} }-\hat{p}_{2k+1}^{i_{2},\omega _{c} },
\end{aligned}
\end{equation}
$\hat{x}_{k}^{s_{1},\omega _{a} }$ denotes the quadrature amplitude operator of the signal field with frequency $\omega_{a}$ from NOPA1. $s_1$ and $i_1$ represent the signal and idler fields from NOPA1. $s_2$ and $i_2$ represent the signal and idler fields from NOPA2.

Similarly, considering the VLF criterion for a minimum unit of the 3D cluster state, the corresponding inequalities can be obtained based on eight basic modes and 127 possible patterns,
\begin{equation}
\left \langle \left (\hat{n}_{2k}^{x1}\right )^2\right \rangle<\hbar\qquad       \left \langle \left (\hat{n}_{2k}^{p1} \right )^2\right \rangle<\hbar\qquad for\  all\  k,
\end{equation}
which can be further expressed as:
\begin{equation}
\begin{aligned}
\left \langle \left ( \hat{x}_{2k}^{s_{1},\omega _{a} }+\hat{x}_{2k}^{i_{1},\omega _{a} }+\hat{x}_{2k}^{s_{2},\omega _{a} }+\hat{x}_{2k}^{i_{2},\omega _{a} }-\hat{x}_{2k+1}^{s_{1},\omega _{b} }+\hat{x}_{2k+1}^{i_{1},\omega _{b} }-\hat{x}_{2k+1}^{s_{2},\omega _{b} }+\hat{x}_{2k+1}^{i_{2},\omega _{b} }\right )^2\right\rangle=4e^{-2r_{s1}}< 1,\\
\left \langle \left ( -\hat{p}_{2k}^{s_{1},\omega _{a} }-\hat{p}_{2k}^{i_{1},\omega _{a} }-\hat{p}_{2k}^{s_{2},\omega _{a} }-\hat{p}_{2k}^{i_{2},\omega _{a} }-\hat{p}_{2k+1}^{s_{1},\omega _{b} }+\hat{p}_{2k+1}^{i_{1},\omega _{b} }-\hat{p}_{2k+1}^{s_{2},\omega _{b} }+\hat{p}_{2k+1}^{i_{2},\omega _{b} }\right )^2\right\rangle=4e^{-2r_{i1}}< 1.
\end{aligned}
\end{equation}

As is shown in Fig.6, the sufficient condition for inseparability of the generated 3D cluster state is satisfied when the squeezing is larger than $-6dB$. Adding one NOPA gives each mode a higher degree of freedom, which result in a bigger required squeezing of the 3D case. 
\begin{figure}[H]
\begin{minipage}{0.49\linewidth}
\vspace{3pt}
\centerline{\includegraphics[width=0.8\textwidth]{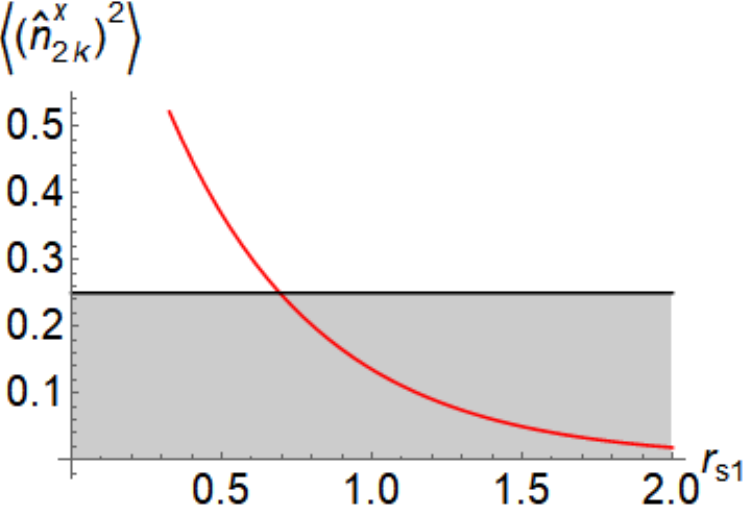}}
\centerline{(a)}
\end{minipage}
\begin{minipage}{0.49\linewidth}
\vspace{3pt}
\centerline{\includegraphics[width=0.8\textwidth]{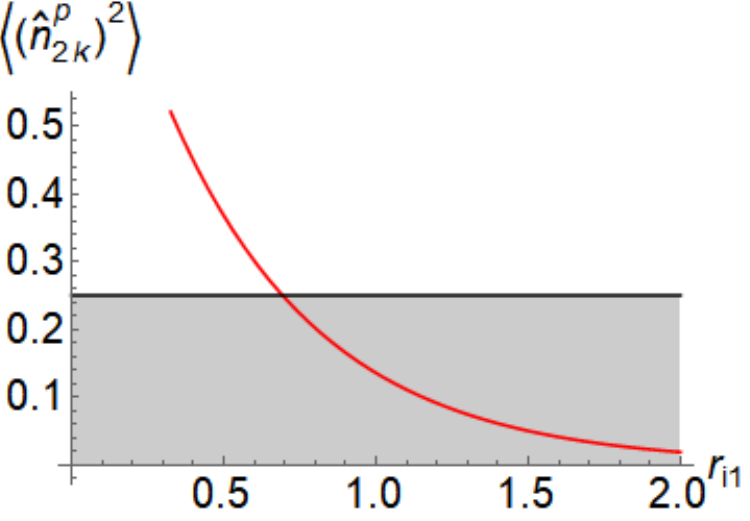}}
\centerline{(b)}
\end{minipage}
\caption{The nullifier of quadrature $\hat{x}$ (a) and $\hat{p}$ (b) versus the squeezing parameter, for the generated 3D cluster state. The squeezing at the intersection is about $-6dB$.}
\end{figure}

\section{Quantum computing based on the generated cluster states}
\label{sec:5}

On the one hand, the generated 1D dual-rail cluster states are necessary resources for MBQC, and can be used to achieve single-mode MBQC by applying the concatenation of teleportation-based input-coupling scheme, and not have to erase a half or three fourths of the generated cluster states before using\cite{menicucci(2011),yokoyama(2013)}. A possible four-step quantum computing circuit utilizing this scheme is shown in Fig.7(a). The input state $\left | \psi_{in}\right \rangle$ and the signal mode $S_1$ (comes from NOPA at $t_1$ time) are coupled via BS1, then two observables $\hat{a}_{in}\left(\theta _{in}\right)$ and $\hat{a}_{S1}  \left(\theta _{S1} \right)$ ($ \hat{a}\left ( \theta\right ) =\left ( \hat{x} \cos \theta +\hat{p}\sin \theta   \right )$) can be measured, thus the first quantum teleportation is completed. After that, the idler mode $I_1$ becomes the new input state and is coupled with the signal mode $S_2$ on BS2, and the corresponding observables can be measured...... in the same way, by repeating couplings and measurement operations, all steps of the four-step teleportation can be completed, and all measured results can contribute to the final output state through the corresponding feedforward operator $\hat{X}$ and $\hat{Z}$, where $\hat{X}\left(s\right )=e^{-2is\hat{p}}$ and $\hat{Z}\left(s\right )=e^{2is\hat{x}}$ ($s$ is the measurement value) are the position and momentum displacement operators. The resulting output state $\left|\psi_{out}\right \rangle$ can be expressed as:
\begin{equation}
\left|\psi_{out}\right \rangle=M\left ( \theta _{+4},\theta _{-4}\right )M\left ( \theta _{+3},\theta _{-3}\right )M\left ( \theta _{+2},\theta _{-2}\right )M\left ( \theta _{+1},\theta _{-1}\right )\left|\psi_{in}\right \rangle,
\end{equation}
where $M\left(\theta _{+1},\theta _{-1}\right)$ represents the transformation relation of the first teleportation. By taking $\theta _{-1}=\frac{\pi}{2}$, the transformation of the two-step teleportation is
\begin{align}
M\left ( \theta _{+2},\theta _{-2}\right )M\left ( \theta _{+1},\theta _{-1}\right )=R\left (-\frac{\pi}{2}-\frac{1}{2}\theta _{+2}\right )S\left (\ln{\tan \frac{1}{2}\theta _{-2} }\right )R\left (-\frac{\pi}{2}-\frac{1}{2}\theta _{+2}-\theta _{+1}\right ),
\end{align}
where $\theta_{\pm1}=\theta_{in}\pm\theta_{S_1}$, here $\theta _{in}$ and $\theta _{S_1}$ represent the relative phases between the signal and local beams in homodyne measurements of modes $in$ and $S_1$, and then all other $\theta_{\pm k},k=2,3,4$, can be determined in the same way. Based on the Bloch-Messiah reduction\cite{braunstein(2005)}, an arbitrary single-mode Gaussian operation (such as rotation, squeezing and shearing operations) can be decomposed into $R\left ( \theta _{2}\right )S\left (r \right )R\left (\theta _{1}\right )$, thus two-step teleportation is enough to obtain an arbitrary single-mode Gaussian operation, as is shown in Eq.(18). Furthermore, teleportation of more steps is more effective in realizing more single-mode Gaussian operations in succession. Specially, in our scheme, many 1D dual-rail cluster states are generated parallelly, which means different single-mode Gaussian operations can be implemented simultaneously.
\begin{figure}[H]
\begin{minipage}{0.49\linewidth}
\vspace{3pt}
\centerline{\includegraphics[width=1.2\textwidth]{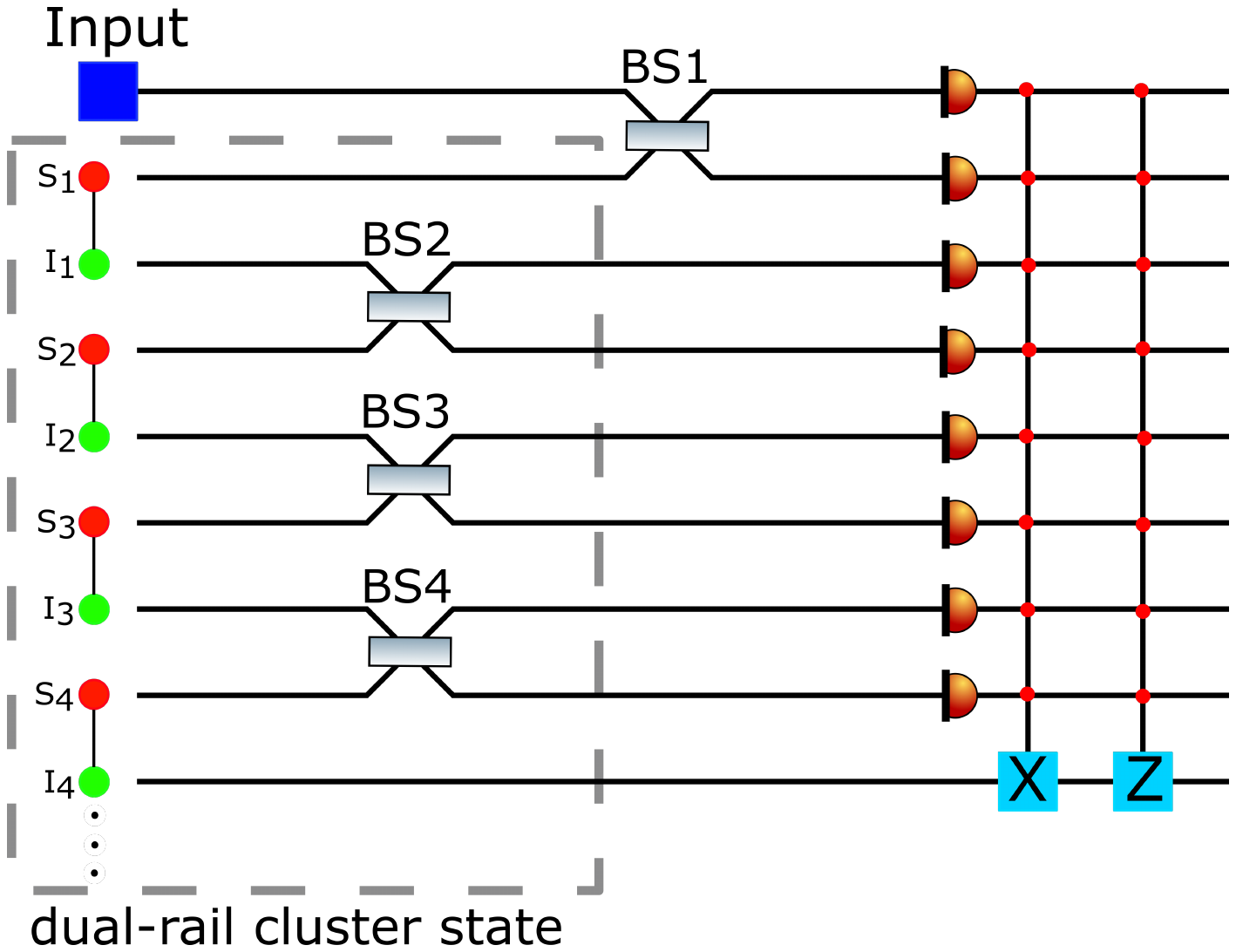}}
\centerline{(a)}
\end{minipage}
\begin{minipage}{0.49\linewidth}
\vspace{3pt}
\centerline{\includegraphics[width=1.2\textwidth]{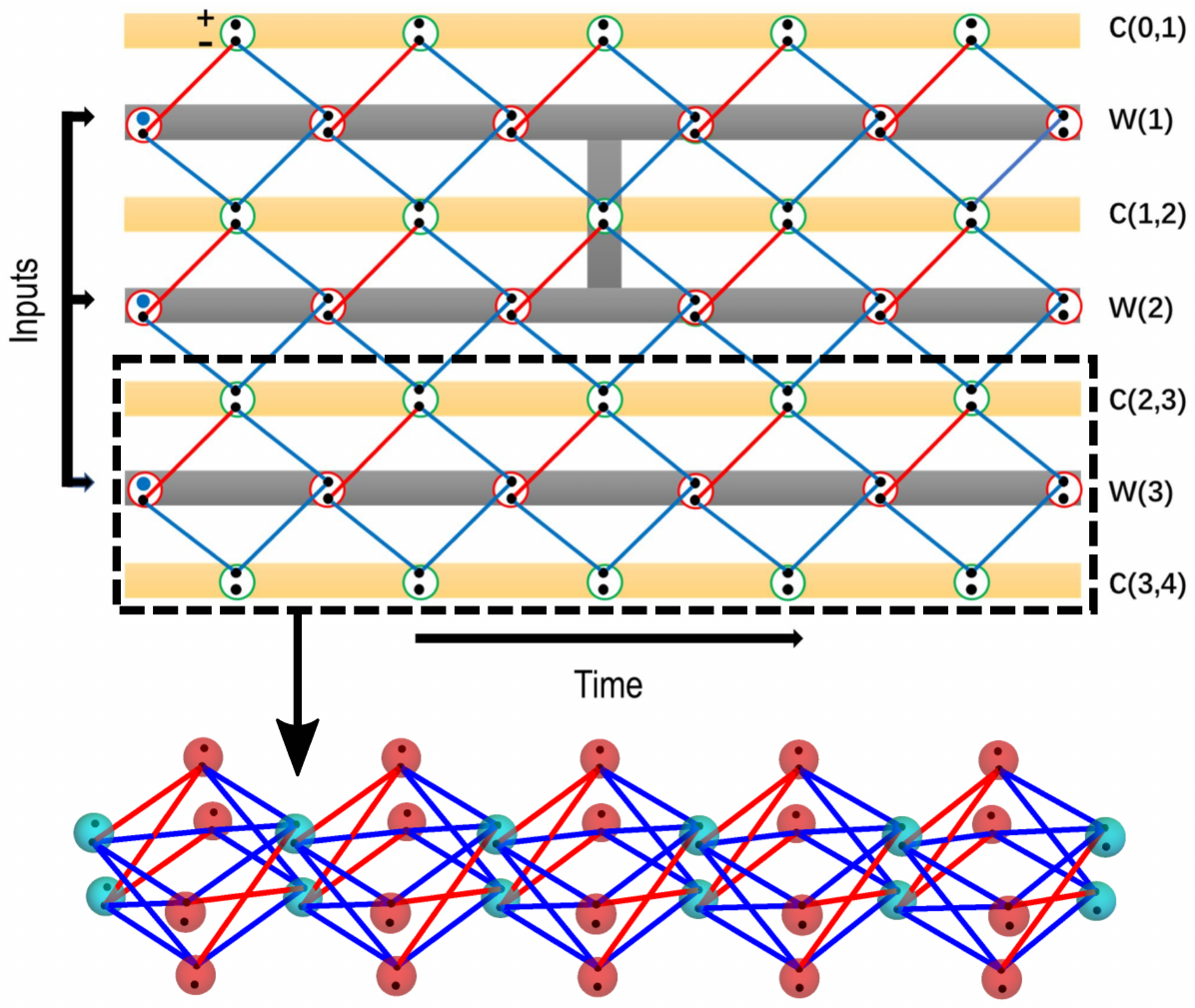}}
\centerline{(b)}
\end{minipage}
\caption{Demonstration of measurement-based quantum computing scheme using temporal-mode 1D dual-rail (a) and 3D (b) cluster states. }
\end{figure}

On the other hand, the generated 3D cluster state can achieve universal MBQC by using the macronode protocol\cite{alexander(2014),alexander(2016),wu(2020)}. The nodes (red and aqua ball) in Fig.4(a) which contain two physical modes $S$ (signal) and $I$ (idler), are considered as macronodes. Each given macronode  $\left\{ S,I\right\}$ can be mapped into a distinct distributed mode $\left\{ +,-\right\}$, 
\begin{equation}
\hat{a}_{\pm}: =\frac{1}{\sqrt{2} }\left( \hat{a}_{S}\pm \hat{a}_{I}\right). 
\end{equation}

As is shown in Fig.7(b), each layer of the generated 3D cluster state can be simplified to a graph of numerous disjoint squares by expressing it in terms of distributed-modes, which deﬁnes a non-local tensor product structure for each macronode. Red and green macronodes represent wire (W(i)) and control (C(i,i+1)) macronodes. When local measurement is implemented at wire (control) macronodes with $\theta _{Sm} \ne \theta _{Im}$, the corresponding square subgraphs will connect with their neighbors in horizontal (vertical) direction, otherwise ( $\theta _{Sm} = \theta _{Im}$) they won't connect. Therefore, local measurements can control the connection condition. For instance, measurements on C(i, i +1) can control w(i) and w(i +1) connected or disconnected. In Fig.7(b), the quantum information can be encoded in the blue modes and flows along each gray channel, sometime it can be transferred among different channels under the control of orange operations. 

Moreover, the 3D cluster state generated from our scheme has two layers, which means it is easier to realize more single-mode and two-mode Gaussian quantum operations and it also has more advantages in fault-tolerant and error correction of MBQC\cite{vuillot(2019),noh2020,fukui(2018)}.

\section{Conclusion}
\label{sec:6}

Schemes of generating 1D and 3D CV cluster states by multiplexing in both frequency and time domain are proposed. The generated 1D cluster states correspond to dual-rail cluster states, while the generated 3D cluster states is a bilayer square-lattice structured entangled state and have special applications in quantum computing. In the time and frequency domain, the number of obtained entangled modes can reach $10^4\sim10^6$\cite{yoshikawa(2016)}, and $10^2\sim10^4$\cite{PWang(2014)}, respectively, under present experimental condition. Therefore, the number of entangled modes in our scheme can reach $10^{6}\sim10^{10}$, which is ultra-large scale. In addition, the squeezing levels that are required to build the 1D and 3D cluster states are $-3dB$ and $-6dB$, respectively. This means that our schemes are experimental feasible, due to to developed techniques of optical frequency comb and squeezed state generation\cite{Vahlbruch(2016)}.

On the other hand, we discuss the concrete applications of the generated large-scale 1D dual-rail and 3D bilayer CV cluster states for MBQC. The generated 1D and 3D cluster states can achieve single-mode and universal MBQC\cite{Asavanant(2021)}, respectively, and every degree of freedom can be fully utilised. Specially, the capability of producing this kind of 3D cluster state is of great significant, due to the requirements of performing  efficient coding and topological error correction\cite{Larsen(2021)}. When further combing with non-Gaussian Gottesman-Kitaev-Preskill coding\cite{vuillot(2019),noh2020,fukui(2018)}, it will make the fault-tolerant quantum computing possible.

\begin{backmatter}
\bmsection{Funding}
Natural Science Foundation of China (NSFC) (11874249, 11874248,11974225,12074233); National Key Research and Development Program of China (2021YFA1402002, 2021YFC2201802).

\bmsection{Disclosures}
The authors declare no conflicts of interest.

\bmsection{Data availability} Data underlying the results presented in this paper are not publicly available at this time but may be obtained from the authors upon reasonable request.
\end{backmatter}


\bibliography{mainref}
\end{document}